# Uncovering domain morphology in an unconventional magnet with scanning diamond quantum magnetometry

Freya Johnson[1], Jan Zemen[2], Helena Knowles[1], Lesley F. Cohen[3]

[1] Cavendish Laboratory, University of Cambridge, Cambridge, CB3 0HE UK

[2] Faculty of Electrical Engineering, Czech Technical University in Prague, Technická 2, Prague, 160 00 Praha 6 Czech Republic

[3] Department of Physics, Blackett Laboratory, Imperial College London, London SW72AZ, UK

**Abstract**

Unconventional magnetic materials including non-collinear antiferromagnets, p-wave magnets and altermagnets, are an emerging frontier for quantum spintronics and hybrid quantum devices. Critical to the application of these materials is control over the magnetic domain state, as their unique, symmetry-driven properties vanish in a multi-domain limit. However, the mechanisms governing domain formation in materials with compensated local moments remain poorly understood. In this work, we examine the ferrimagnetic to non-collinear antiferromagnetic phase transition of $Mn_3NiN$ using scanning nitrogen-vacancy centre magnetometry. We provide nanoscale mapping of the magnetic domain evolution on cooling and correlate the local stray fields with global magnetometry and anomalous Hall effect measurements. We observe the formation of a disordered, dendritic domain structure whose roughness is quantified using its fractal dimension. The fractal dimension steadily increases on cooling through the transition, saturating at a value of ~ 1.55 in the non-collinear phase, but the domain area distribution does not show any significant changes. We show this behaviour cannot be explained by the balance of demagnetisation energy and domain wall energy, and conclude elastic contributions and defects are a critical factor to explain the domain size.

**Introduction**

Integrating magnetic materials with superconductors has been shown to lead to a plethora of novel and interesting phenomena [1, 2]. In particular, this combination can realise triplet superconductivity, which has the potential to enable quantum computing with ultra-low power cryogenic memory exploiting dissipationless supercurrents [3, 4] and is the key to unlock topological quantum computing [5, 6]. However, using conventional ferromagnetic materials is challenging due to their finite magnetisation, which causes parasitic stray fields. These lead to local suppression of superconductivity, which is exacerbated in a nanoscale device, negatively impacting scalability [7, 8]. For this reason, attention has turned to integrating unconventional magnetic materials with superconductors [8]. These materials demonstrate vanishing net magnetisation, but possess an anisotropic spin-polarised Fermi surface due to the particular symmetries of their crystal and magnetic structure, and include planar non-collinear antiferromagnets (nc-AFM) [9], d-wave altermagnets [10] and non-coplanar p-wave magnets [11].

Non-collinear antiferromagnets have already been shown to possess many different enhanced functional properties - such as anomalous Hall effect [12-14], magneto-optical Kerr effect [15, 16] and anomalous Nernst effect [17-19] due to the presence of Berry curvature in the band structure, allowing for easy read-out of the magnetic state. The interaction between these fascinating materials and superconductivity has also been shown to lead to the generation of unexpectedly large triplet supercurrents [20], and has been integrated into d.c. SQUIDs [21]. However, the fundamental symmetry driven highly anisotropic properties of these materials vanish in a multidomain state. For this reason, understanding the factors that drive domain formation and structure in these novel materials is of the utmost importance for technological applications.

In this work, we investigate these mechanisms in the nc-AFM $Mn_3NiN$. When grown as a thin film with compressive strain in-plane imparted by a suitable substrate, in this case $SrTiO_3$ (STO), the strain stabilises a ferrimagnetic phase above its Néel temperature, $T_N$ [22]. This provides an ideal environment to examine the influence of magnetisation and magnetic order on the domain structure in both phases and across the transition. We have previously examined the domain structure in a Hall bar of this material using scanning anomalous Nernst imaging, revealing a checkerboard pattern of domains with sizes ~1 μm in area, limited to the spot size of our scanning laser [17]. We build upon this work by employing scanning nitrogen-vacancy centre quantum magnetometry (NV) on an unpatterned sample. This technique has emerged as a sensitive, high-resolution non-invasive probe of ferromagnetic and antiferromagnetic order, particularly for thin film samples [23, 24] and 2D materials [25]. Here, we provide the first direct, nanoscale visualization of the domain texture across the magnetic phase transition in $Mn_3NiN$.

**Methods**

30 nm thick $Mn_3NiN$ films were deposited using pulsed laser deposition at 400 °C on single crystal (001)-oriented $SrTiO_3$ (STO) substrates, as described in detail Ref. [26].

Magnetometry was performed using a Quantum Design Superconducting Quantum Interference Device magnetometer. Magnetotransport data were collected using the van der Pauw method and the resistivity option of Quantum Design Physical Properties Measurement system. X-ray reflectivity was measured using a Malvern Analytical Empyrean MultiCore High-Performance X-ray Diffractometer.

The scanning nitrogen-vacancy centre magnetometry setup is described in detail in the supplementary information of Ref. [24]. In brief, a cryogenic atomic force microscope system scans a diamond sensor ~ 50 nm above the sample. Microwave excitation is applied via a copper wire stretched over the sample. Pulsed-optically detected magnetic resonance is employed to determine the on-axis field magnitude.

**Results**

In both the ferrimagnetic phase and the nc-AFM phase of $Mn_3NiN$, spin canting due to strain results in a small net moment in the <112> type directions – the piezomagnetic effect. [26-28] This effect produces the small stray fields detected using our NV probe. Figure 1a show the schematic of the experimental set-up. A diamond probe with [111] orientation is scanned above the sample surface in an applied magnetic field of 150 mT, simultaneously measuring the topography (Figure 1c) and the stray field (Figure 1d) on cooling. We first characterise the sample using X-ray reflectivity, from which we extract the thickness of the film as 32 nm and an interfacial roughness of 1 nm. Then, we perform scanning NV measurements at 260 K, when the magnetic order is ferrimagnetic. The surface topography is smooth, with a root mean squared roughness of 0.48 nm, other than a low density of particulates or droplets common in growth using PLD. The corresponding magnetic field map reveals broad, smooth domains with a weak field magnitude varying between ~ ± 60 mT and whose structure appears uncorrelated with the droplets. We mask off these features and perform a pixel-wise correlation between the topography and stray field. The resulting analysis (Figure 1e) yields a Pearson coefficient r = -0.03, confirming the absence of correlation between the two images and demonstrating the probe is genuinely sensitive to the magnetic landscape rather than to surface-related artefacts.

Next, we cool to 190 K in a field of +150 mT applied out-of-plane, taking further images on cooling. Due to drift in the cooling procedure, each image is taken from a different section of the sample surface. Representative images are shown in Figure 2a-c. With the high spatial resolution of scanning NV, we are able to identify multiple domains in our 4 µm$^2$ scan. The magnitude of the stray fields from the sample increases on cooling to +/- 200 µT, which agrees with other measurements of antiferromagnetic materials using this technique. [23, 29, 30] Qualitatively, the domains appear more irregular and interconnected as the sample cools.

To quantify these changes, two spatially averaged metrics were extracted from each stray field map: the mean field and the mean absolute field. While these metrics average out local domain structure, they enable a direct comparison with global measurements of magnetization and the anomalous Hall effect. In Figure 3a the mean field per micron of each image is plotted as a function of temperature and overlaid with field-cooled cooling (FCC) magnetization data, which was acquired under a +150 mT out-of-plane field. From the peak in FCC, we identify $T_N$ = 210 K, which is consistent with previous reports [22]. The mean field follows a similar temperature dependence. Unexpectedly, a negative mean field is observed at 260 K and 250 K. Given that the local field is calibrated via the Zeeman shift measured at a reference point far from the sample, we attribute this apparent negative value to a small systematic offset in the field calibration.

To measure the anomalous Hall resistivity $\rho_H$, transverse resistivity $\rho_{xy}$ is measured on cooling in +/- 150 mT and the results are antisymmetrised according to $\rho_H = (\rho_{xy}(+B) - \rho_{xy}(-B)) / 2$. As is shown in Figure 3b, $\rho_H$ is present at all temperatures, reaching its maximum value at $T_N$ and falling in the nc-AFM phase, which is also in agreement with previous work. [15] In contrast, the spatially averaged mean absolute field (Figure 3c), which serves as a proxy for the local magnetization magnitude, rises steadily and saturates in the nc-AFM phase. This is a clear demonstration that the AHE in this material's nc-AFM phase is not a simple function of its magnetization. [31]

Having established that our NV images are consistent with global measurements and uncorrelated with topography, we can confidently attribute the observed stray fields to the intrinsic magnetic structure of the film. We now turn to a quantitative analysis of the domain geometry. In thin films with perpendicular magnetic anisotropy (PMA), the fractal dimension, $D_f$, is a useful parameter for characterizing domain "jaggedness," even if the structures are not strictly fractal. [32-34] We therefore examined the evolution of $D_f$ across the ferrimagnetic to nc-AFM transition. To achieve this, the stray field maps were first binarized using Otsu's thresholding method to segment the positive and negative domains. This process is illustrated for the 190 K map in Figure 4a-d. The perimeter and area of each domain were then extracted and plotted on a log-log scale (Figure 4e). The fractal dimension was extracted from the slope of this plot according to the scaling relation $P \propto A^{D_f / 2}$. [32] The temperature dependence of $D_f$ is shown in Figure 4f. On cooling, we find $D_f$ steadily increases from 1.39 in the ferrimagnetic phase, saturating at 1.56 in the nc-AFM phase. This behaviour mirrors the trend observed in the mean absolute field. Indeed, plotting the fractal dimension versus the mean absolute field reveals a clear linear relationship (Figure 4g), indicating that the increasing magnetic strength is associated with the development of more complex, jagged domain structures.

**Discussion**

Our observations suggest that the antiferromagnetic domains in the non-collinear phase are first nucleated in the ferrimagnetic phase, grow in complexity on cooling, and ultimately form a complex, labyrinthine domain structure. This increase in domain complexity is reminiscent of the behaviour of PMA thin films, where $D_f$ has been found to increase with saturation magnetisation in PMA Co/Pd multilayers. [32] The formation of these structures have been described by considering the interaction between randomly distributed pinning sites and the domain wall stiffness. [33] When the domain wall stiffness is small, it may more easily be deformed by pinning, leading to the observed jagged growth and higher fractal dimension. [33] However, the domain wall stiffness itself is determined by the balance between the demagnetisation energy, which favours domain formation, and the domain wall energy, which penalizes it. [35, 36].

Although we observe $D_f$ increases as mean absolute field increases, when we consider the relevant energy scales in $Mn_3NiN$ we conclude that they must be independent. As the demagnetisation energy scales with the square of the magnetization, it is significantly weaker in this system than in a typical ferromagnet. Our scanning NV measurements confirm this, showing maximum stray fields of approximately 0.2 mT, an order of magnitude smaller than the fields observed above a PMA thin film of CoFeB, [37] and therefore we expect the demagnetisation energy to be ~100x smaller. However, we must also consider the domain wall energy. Using atomistic simulations, we have previously obtained a domain wall energy of 1.6 mJ/m$^2$ in nc-AFM $Mn_3NiN$ at zero temperature – the energy at 190 K will be lower as $\sigma \propto \sqrt{AK_u}$ where the exchange stiffness $A$ and the uniaxial anisotropy constant $K_u$ both decrease with increasing temperature. [35] As a comparison, the domain wall energy in CoFeB has been reported in the range 4-8 mJ/m$^2$ at 100 K. So, while $Mn_3NiN$ likely has ~ 5x smaller domain wall energy than CoFeB, the difference is not enough to counteract the effect of the vanishing demagnetisation, and we should expect high domain wall stiffness and large smooth domains with small $D_f$ – the opposite of what we see experimentally.

For this reason, we now must review the role of defects and strain. Defects can create regions of local strain within the film that induce a local moment via the piezomagnetic effect [38], increase the local anisotropy, and change the strength of exchange interactions, which in turn changes the domain wall energy and leads to domain wall nucleation or pinning. Moreover, we can also consider the balance of domain wall energy and magnetoelastic energy. If we first consider the case of no defects, the global biaxial in-plane strain ($\varepsilon_{xx}= \varepsilon_{yy}$ ~-0.2% [22]) due to lattice mismatch of $Mn_3NiN$ film with STO substrate would induce uniform magnetization of ~0.05 $\mu_B$/f.u. [28]. However, the local defects and related uniaxial in-plane strains ($\varepsilon_{xx}\neq\varepsilon_{yy}$ ~1%) lead to much larger net magnetization of ~0.25 $\mu_B$/f.u. [22, 38]. The large local magnetization at defects may polarize moments in surrounding defect-free regions. The enhanced magnetization in this region is not proportional to the small background strain induced by the substrate, so it will add to the elastic energy of the system owing to the inverse piezomagnetic effect. Such extra magnetoelastic energy can be reduced by mutual compensation of contributions from domains with different direction of magnetization. The magnetostrictive version of this effect has been invoked in the collinear antiferromagnets CuMnAs and Mn$_2$Au to explain the domain structures at the edges of patterned devices [39].

Our previous transmission electron microscopy (TEM) studies on Mn$_3$NiN/STO films confirm the presence of a high density of edge dislocation defects extending approximately 20 nm from the film-substrate interface, and a global strain due to mismatch with the substrate [38]. While the limited field of view in TEM (~35 nm$^2$) precludes a direct one-to-one comparison with the micron-scale NV maps, these snapshots revealed an inhomogeneous strain landscape. A detailed model of domain morphology requires atomistic simulations of the spin structure in the vicinity of the slip-planes which is beyond the scope of this study. However, we can estimate the average elastic energy cost due to the excess magnetization and resulting piezomagnetic strain and compare to domain wall energy to give an estimate of the average domain length scale, $L$ using the theory of Kittel [40]. We begin by fitting the exchange constants of the Heisenberg model, $J_{ij}$, to energies calculated from density functional theory [28]:

$$H = -\sum_{i \neq j} J_{ij} \boldsymbol{e}_i \cdot \boldsymbol{e}_j \quad (1)$$

Where $e_i$ is the spin on site $i$. We obtain $J_{ij}$ ~ -300 meV/f.u. for all nearest neighbour pairs in the unstrained case. These parameters are then used to calculate the spin-wave stiffness, $D$, and from this the exchange stiffness, $A$ [41]:

$$D = \frac{2\mu_B}{3M} \sum_j J_{0j} R_{0j}^2 \quad (2)$$

$$A = \frac{M_s \cdot D}{2g\mu_B} \quad (3)$$

where the summation is restricted to nearest neighbours in our case, $R_{0j}$ are the distances to nearest neighbours, $M$ is the local moment, $M_s$ is saturation magnetization and $g$ and $\mu_B$ are the Landé factor and Bohr magneton. From $A$, and the uniaxial anisotropy constant, $K$, we can estimate the DW width $\delta$ and DW energy, $\gamma$:

$$\delta = \pi\sqrt{A/K} \quad (4)$$

$$\gamma = 4\sqrt{AK} \quad (5)$$

From this, we obtain $\gamma$ = 1.35 mJ / m$^2$, in reasonable agreement with the value of 1.6 mJ/m$^2$ from atomistic simulations. Finally, the average domain length scale $L$ can be obtained from the balance of the magnetic domain wall and elastic contribution to the free energy:

$$E(L) \approx \gamma \frac{t}{L} + \kappa Y \varepsilon_0^2 L \quad (6)$$

$$L = \sqrt{\kappa \frac{\gamma t}{Y \varepsilon_0^2}} \quad (7)$$

where $t$ = 32 nm is the film thickness, $Y$ = 130 GPa is the Young's modulus, $\varepsilon_0$ = 0.1% is the extra strain induced by excess magnetization induced by nearby defects, $M_{exc}$ = 0.02 μ$_B$/f.u. (measure of piezomagnetic effect), and $\kappa$ is a geometric factor that we set to 1. From this, we obtain $L$ of ~ 500 nm. As the underlying defects and strains are static, this model predicts that the equilibrium domain size should not have a strong temperature dependence. We caution that the theory above has been derived for ferromagnetic films with stripe domains. However, we describe our material by $M_{exc}$ corresponding to half the saturation magnetization expected in a defect-free film, to get the induced strain, and by the local moment $M$ = 2.8 $\mu_B$/atom to get the spin-wave stiffness, so our simple estimate captures the presence of the large Mn moments which mostly compensate each other.

We note that the piezomagnetic effect inducing the local magnetization at the defects is expected to be small in the collinear ferrimagnetic phase (where the moments do not cant) and gradually grow during cooling to reach a saturated value in the noncollinear-AFM phase below $T_N$, where the magnetoelastic coupling is fully developed. This will impact the DW energy as well as the elastic energy, which is driven by the inverse piezomagnetic effect. The increase in the magnetoelastic energy can explain the increase in $D_f$ on cooling and saturation of $D_f$ in AFM phase. In Figure 5 we present the measured domain area distribution at each temperature. The median area fluctuates between 0.02 – 0.06 um$^2$ (except the outlier at 250 K), corresponding to $L$ of ~ 140 – 250 nm in reasonable agreement with our estimated size, and there is no clear trend with temperature, as expected in case of high concentration of defects which act as pinning sites.

**Conclusion**

In conclusion, we have employed scanning diamond quantum magnetometry to map magnetic domain textures across the ferrimagnetic to non-collinear antiferromagnetic phase transition in an Mn$_3$NiN thin film. Our images reveal a clear evolution in domain morphology, from smooth boundaries in the ferrimagnetic phase to a complex rough structure at lower temperatures, which we quantify using the domain size distribution and fractal dimension. Our analysis reveals that this characteristic structure cannot be explained by a conventional balance of demagnetisation and domain wall energies; instead, we suggest magnetoelastic contributions play the dominant role. These findings have significant implications for applying unconventional magnets in quantum technologies, as these complex, multi-domain states can negate their useful symmetry-driven properties. However, sensitivity to internal strain also presents an opportunity. Our work suggests that controlling the strain distribution in the films, for example through patterning, choice of substrate or externally applied force provides a pathway towards designing specific domain states. Future work will focus on imaging domains in patterned nanostructures and in annealed samples with lower defect densities to explore this potential for domain engineering.

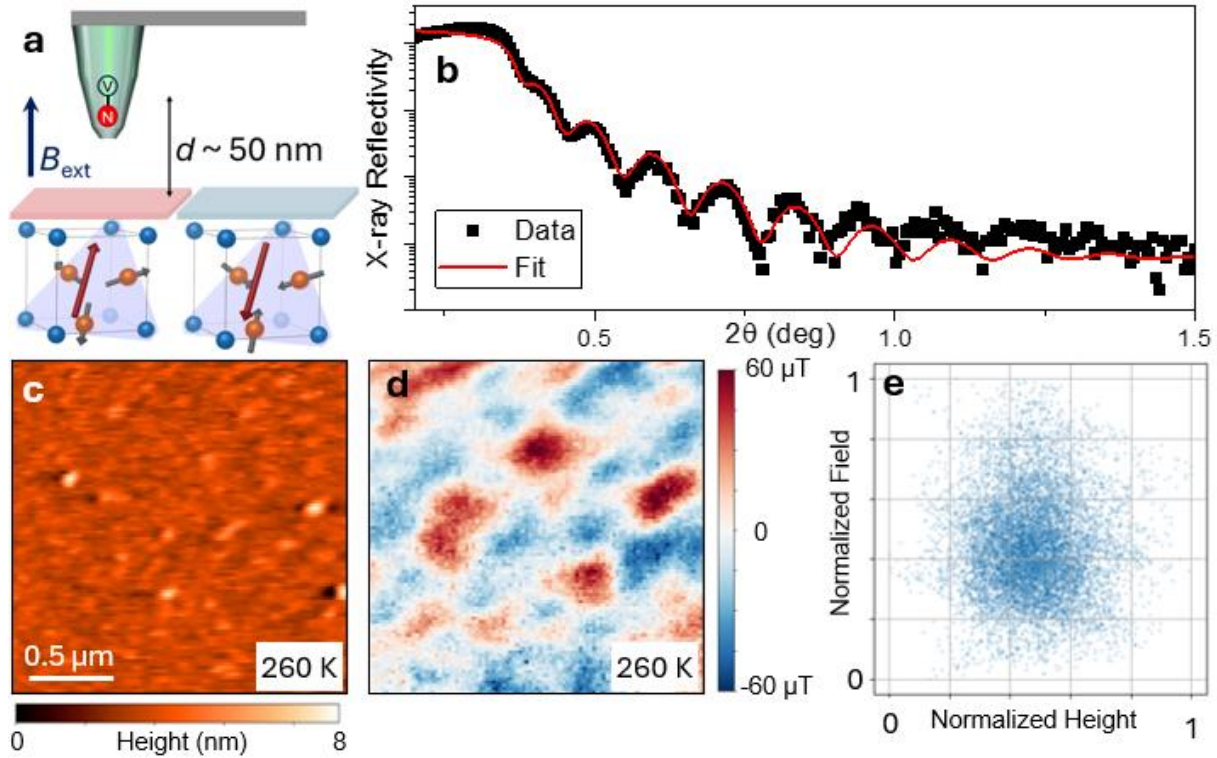

*Figure 1. Mn$_3$NiN sample characterisation and NV imaging in the ferrimagnetic phase. (a) Schematic showing measurement set-up. A diamond tip with [111] orientation is scanned 50 nm above the sample surface. Strain induced canting in the non-collinear structure leads to a weak net moment in the <112> type directions as indicated by the red arrows. b) X-ray reflectivity scan of the sample, from which we extract thickness = 32 ± 1 nm and interfacial roughness = 1 ± 0.3 nm. Through optically detected magnetic resonance, we detect the component of magnetic field strength oriented in the out-of-plane direction (c) Topographical and (d) Stray field map of the film surface at 260 K. (e) Pixel-wise correlation between the topography and the stray field map, showing the two are uncorrelated.*

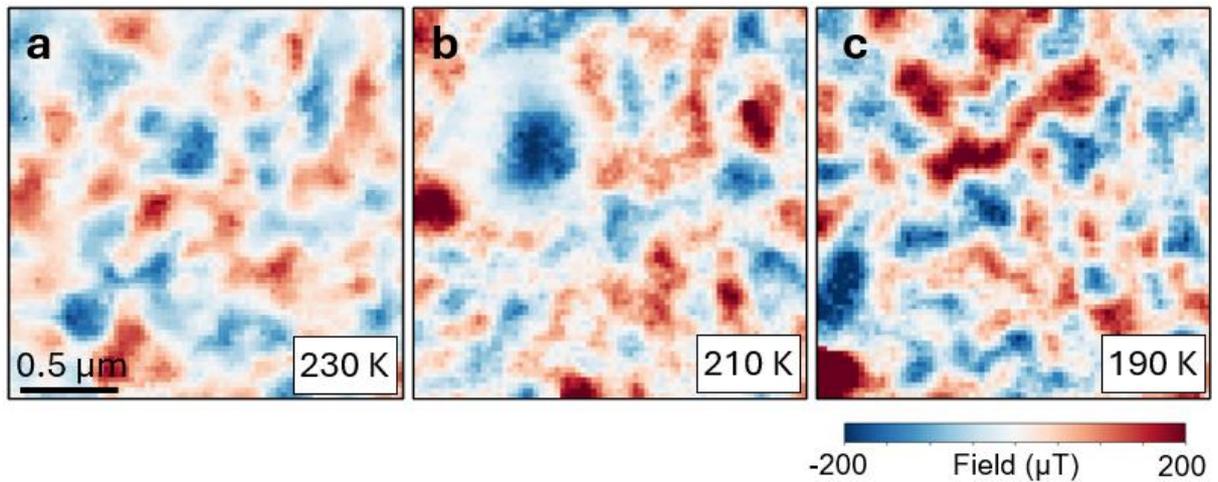

*Figure 2. Scanning nitrogen-vacancy quantum magnetometry images of Mn$_3$NiN on cooling (a) 230 K, (b) 210 K and (c) 190 K.*

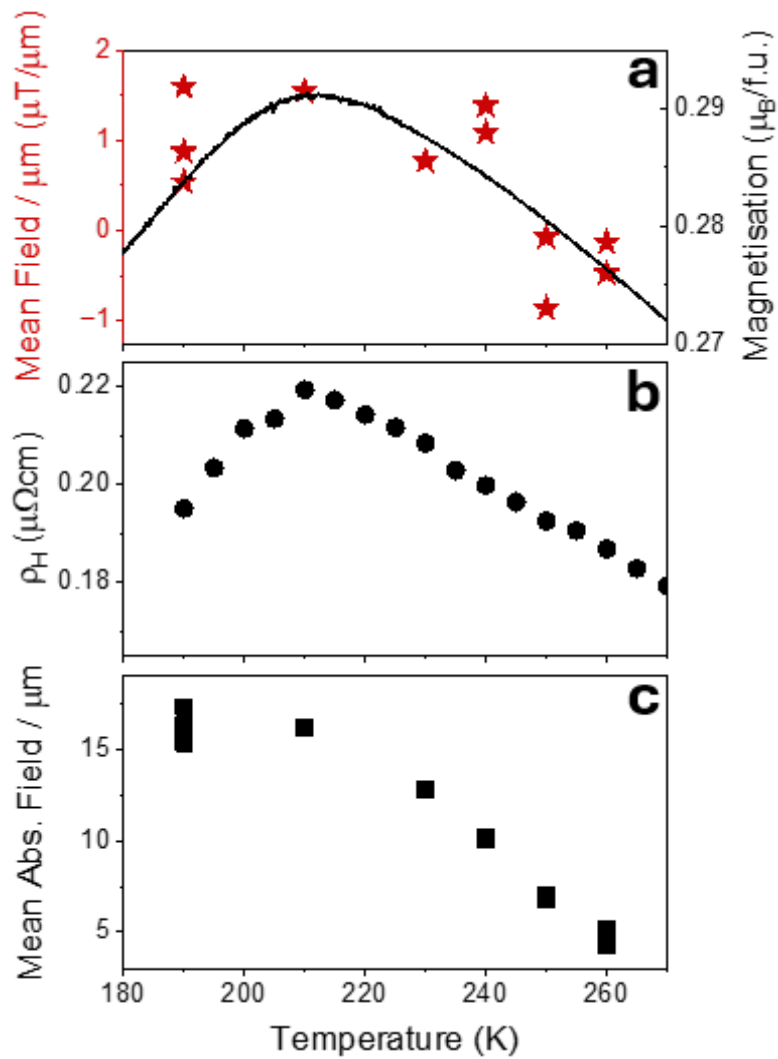

*Figure 3. Comparison between spatially averaged metrics acquired from NV and other global measurements as a function of temperature. (a) Mean field per micron for each image (left axis) and field cooled magnetisation (right axis). (b) Anomalous Hall resistivity, measured on cooling in 150 mT. (c) Mean absolute field per micron.*

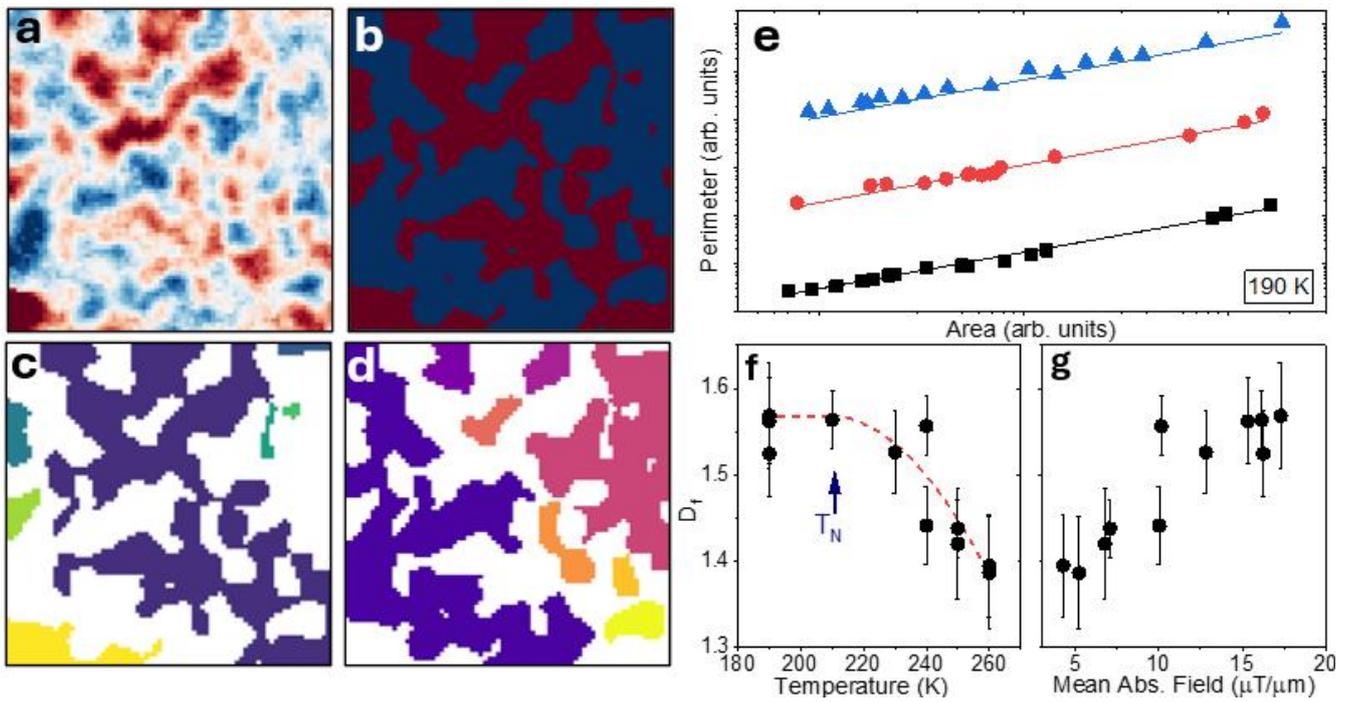

*Figure 4. Domain structure analysis. (a-d) Extracted domain regions. (a) Field map at 190 K. (b) Thresholded image using Otsu's method. (c,d) Regions identified corresponding to (c) positive domains (d) Negative domains. (e) Perimeter vs area of the three images at 190 K, with linear fits from which we extract the fractal dimension. An offset has been added for clarity. (f) Fractal dimension $D_f$ with respect to temperature. The dashed line is a guide to the eye. (g) $D_f$ against mean absolute field.*

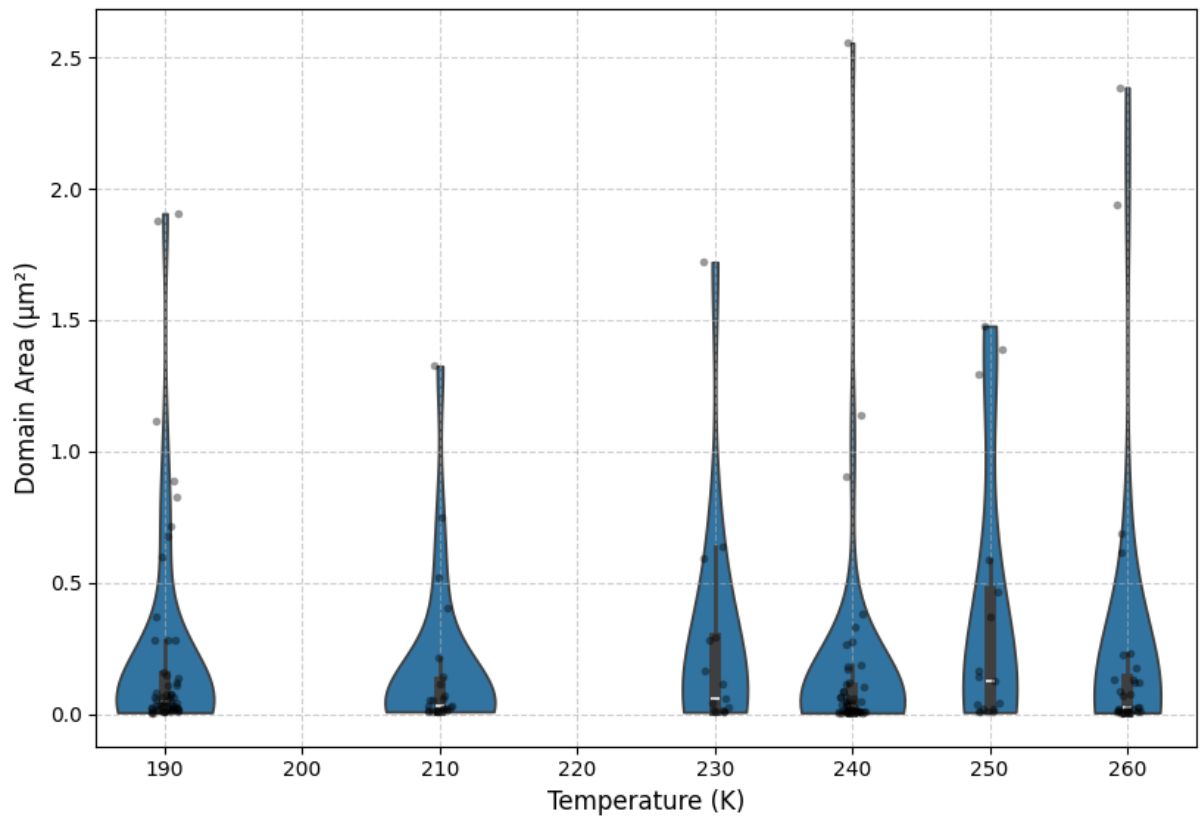

Figure 5. Violin plot showing domain area as a function of temperature.

**References**

1. Bergeret, F.S., A.F. Volkov, and K.B. Efetov, *Odd triplet superconductivity and related phenomena in superconductor-ferromagnet structures.* Reviews of Modern Physics, 2005. **77**(4): p. 1321-1373.
2. Banerjee, N., et al., *Materials for Quantum Technologies: a Roadmap for Spin and Topology.* 2024.
3. Yang, G., C. Ciccarelli, and J.W.A. Robinson, *Boosting spintronics with superconductivity.* APL Materials, 2021. **9**(5).
4. Linder, J. and J.W. Robinson, *Superconducting spintronics.* Nature Physics, 2015. **11**(4): p. 307-315.
5. Sato, M. and Y. Ando, *Topological superconductors: a review.* Reports on Progress in Physics, 2017. **80**(7): p. 076501.
6. Sarma, S.D., M. Freedman, and C. Nayak, *Majorana zero modes and topological quantum computation.* npj Quantum Information, 2015. **1**(1): p. 15001.
7. Birge, N.O. and N. Satchell, *Ferromagnetic materials for Josephson π junctions.* APL Materials, 2024. **12**(4).
8. Fukaya, Y., et al., *Superconducting phenomena in systems with unconventional magnets.* Journal of Physics: Condensed Matter, 2025. **37**(31): p. 313003.
9. Rimmler, B.H., B. Pal, and S.S.P. Parkin, *Non-collinear antiferromagnetic spintronics.* Nature Reviews Materials, 2025. **10**(2): p. 109-127.
10. Šmejkal, L., J. Sinova, and T. Jungwirth, *Emerging Research Landscape of Altermagnetism.* Physical Review X, 2022. **12**(4): p. 040501.
11. Brekke, B., et al., *Minimal models and transport properties of unconventional p-wave magnets.* Phys. Rev. Lett., 2024. **133**.
12. Johnson, F., et al., *Strain dependence of Berry-phase-induced anomalous Hall effect in the non-collinear antiferromagnet Mn3NiN.* Applied Physics Letters, 2021. **119**(22).
13. Nakatsuji, S., N. Kiyohara, and T. Higo, *Large anomalous Hall effect in a non-collinear antiferromagnet at room temperature.* Nature, 2015. **527**(7577): p. 212-215.
14. Boldrin, D., et al., *Anomalous Hall effect in noncollinear antiferromagnetic Mn3NiN thin films.* Physical Review Materials, 2019. **3**(9): p. 094409.
15. Johnson, F., et al., *Room-temperature weak collinear ferrimagnet with symmetry-driven large intrinsic magneto-optic signatures.* Physical Review B, 2023. **107**(1): p. 014404.
16. Higo, T., et al., *Large magneto-optical Kerr effect and imaging of magnetic octupole domains in an antiferromagnetic metal.* Nature Photonics, 2018. **12**(2): p. 73-78.
17. Johnson, F., et al., *Identifying the octupole antiferromagnetic domain orientation in Mn3NiN by scanning anomalous Nernst effect microscopy.* Applied Physics Letters, 2022. **120**(23).
18. Guo, G.-Y. and T.-C. Wang, *Large anomalous Nernst and spin Nernst effects in the noncollinear Mn3X (X=Sn, Ge, Ga).* Physical Review B, 2017. **96**(22): p. 224415.
19. Beckert, S., et al., *Anomalous Nernst effect in Mn3NiN thin films.* Physical Review B, 2023. **108**(2): p. 024420.

20. Jeon, K.-R., et al., *Long-range supercurrents through a chiral non-collinear antiferromagnet in lateral Josephson junctions.* Nature Materials, 2021. **20**(10): p. 1358-1363.
21. Jeon, K.-R., et al., *Chiral antiferromagnetic Josephson junctions as spin-triplet supercurrent spin valves and d.c. SQUIDs.* Nature Nanotechnology, 2023. **18**(7): p. 747-753.
22. Boldrin, D., et al., *The Biaxial Strain Dependence of Magnetic Order in Spin Frustrated Mn3NiN Thin Films.* Advanced Functional Materials, 2019. **29**(40): p. 1902502.
23. Erickson, A., et al., *Nanoscale imaging of antiferromagnetic domains in epitaxial films of Cr2O3 via scanning diamond magnetic probe microscopy.* RSC Advances, 2023. **13**(1): p. 178-185.
24. Tan, A.K.C., et al., *Revealing emergent magnetic charge in an antiferromagnet with diamond quantum magnetometry.* Nature Materials, 2024. **23**(2): p. 205-211.
25. Sun, Q.-C., et al., *Magnetic domains and domain wall pinning in atomically thin CrBr3 revealed by nanoscale imaging.* Nature Communications, 2021. **12**(1): p. 1989.
26. Boldrin, D., et al., *Giant Piezomagnetism in Mn3NiN.* ACS Applied Materials & Interfaces, 2018. **10**(22): p. 18863-18868.
27. Ikhlas, M., et al., *Piezomagnetic switching of the anomalous Hall effect in an antiferromagnet at room temperature.* Nature Physics, 2022. **18**(9): p. 1086-1093.
28. Zemen, J., Z. Gercsi, and K.G. Sandeman, *Piezomagnetism as a counterpart of the magnetovolume effect in magnetically frustrated Mn-based antiperovskite nitrides.* Physical Review B, 2017. **96**(2): p. 024451.
29. Li, S., et al., *Nanoscale Magnetic Domains in Polycrystalline Mn3Sn Films Imaged by a Scanning Single-Spin Magnetometer.* Nano Letters, 2023. **23**(11): p. 5326-5333.
30. Guo, Q., et al., *Current-induced switching of thin film $\ensuremath{\alpha}\text{\ensuremath{-}}{\mathrm{Fe}}_{2}{\mathrm{O}}_{3}$ devices imaged using a scanning single-spin microscope.* Physical Review Materials, 2023. **7**(6): p. 064402.
31. Chen, H., Q. Niu, and A.H. MacDonald, *Anomalous Hall Effect Arising from Noncollinear Antiferromagnetism.* Physical Review Letters, 2014. **112**(1): p. 017205.
32. Kim, D.-H., et al., *Correlation between fractal dimension and reversal behavior of magnetic domain in Co/Pd nanomultilayers.* Applied Physics Letters, 2003. **82**(21): p. 3698-3700.
33. Lyberatos, A., J. Earl, and R.W. Chantrell, *Model of thermally activated magnetization reversal in thin films of amorphous rare-earth-transition-metal alloys.* Physical Review B, 1996. **53**(9): p. 5493-5504.
34. Sayko, G.V., et al., *Fractal domain structures in thin amorphous films.* IEEE Transactions on Magnetics, 1992. **28**(5): p. 2931-2933.
35. Shieh, H.P.D. and M.H. Kryder, *Dynamics and factors controlling regularity of thermomagnetically written domains.* Journal of Applied Physics, 1987. **61**(3): p. 1108-1122.

36. Thiele, A.A., *The Theory of Cylindrical Magnetic Domains.* Bell System Technical Journal, 1969. **48**(10): p. 3287-3335.
37. Rohner, D., et al., *(111)-oriented, single crystal diamond tips for nanoscale scanning probe imaging of out-of-plane magnetic fields.* Applied Physics Letters, 2019. **115**(19).
38. Johnson, F., et al., *The Impact of Local Strain Fields in Noncollinear Antiferromagnetic Films.* Advanced Materials, 2024. **36**(27): p. 2401180.
39. Reimers, S., et al., *Magnetic domain engineering in antiferromagnetic CuMnAs and Mn2Au.* Physical Review Applied, 2024. **21**(6): p. 064030.
40. Kittel, C., *Physical Theory of Ferromagnetic Domains.* Reviews of Modern Physics, 1949. **21**(4): p. 541-583.
41. Pajda, M., et al., *Ab initio calculations of exchange interactions, spin-wave stiffness constants, and Curie temperatures of Fe, Co, and Ni.* Physical Review B, 2001. **64**(17): p. 174402.